\documentclass[letterpaper,11pt]{article}

\usepackage{appendix}
\usepackage{amsmath,amssymb, amsthm}
\usepackage{array}
\usepackage{caption}
\usepackage{cite}
\usepackage{color}
\usepackage{float}
\usepackage{graphicx}
\usepackage{lineno}
\usepackage{mathrsfs}
\usepackage{multirow}
\usepackage{subcaption}
\usepackage{soul}
\usepackage{tabularx}
\usepackage{titlesec}

\titleformat{\subsection}[runin] 
  {\bfseries} 
  {\thesubsection} 
  {0.5em} 
  {} 

\newcommand{\suppress}[1]{}

\begin{document}
\vskip 25mm
\begin{center}

{\large\bfseries  

Competition and survival in modern academia: A  bibliometric case study of theoretical high-energy physics

}

\vskip 6ex

Jarl \textsc{Sidelmann}\footnote{email: \texttt{jarlsidelmann@gmail.com}} \&
Jesper M\o ller \textsc{Grimstrup}\footnote{email: \texttt{jesper.grimstrup@gmail.com}}\\ 
\vskip 6ex

\vskip 3ex

{\footnotesize\it This work is financially supported by entrepreneur Kasper Bloch Gevaldig,\\ Denmark, and by entrepreneur Jeff Cordova, USA}

\end{center}

\vskip 6ex

\begin{abstract}

\noindent We study the career lengths of researchers in theoretical high-energy physics from 1950 to 2020. 
Using a cohort-based analysis and bibliometric data from 30,149 authors in three physics disciplines we observe a dramatic increase in the ratio of academic dropouts over time. While nearly all authors prior to the 1980s remained active more than a decade after their initial publication, this was the case for 50\% or less by the 2010s. Consequently, the fraction of theoretical physicists who achieve full academic careers have diminished significantly in the last four decades. Additionally, we study correlations between author survivability and potential success factors, inferring that early career productivity and collaborative efforts have become increasingly strong determinants over time. A logistic regression is employed to interpret the empirical findings through a survival model. Performance measures further indicate that early career
collaboration and productivity rates have emerged as reliable predictors of author survivability during the period where competition has intensified. Finally, we discuss the potential impacts that these developments may have on science itself.

\end{abstract}

\newpage
\tableofcontents


\section{Introduction}

Competition has always been an integral part of academia. There is competition for funding, for recognition of ideas, and for academic status, but what about competition for survival? How has competition for survival evolved historically and how has it impacted the scientific community? In this paper we attempt to answer these questions.  


An abundance of studies exist analyzing various aspects of success in science (see, e.g., \cite{Wang_Barabási_2021,Santo} and references therein). Most of these focus on the correlation between success and citation counts \cite{Sinatra2,LiuL,Xiaomei, Peters},  productivity \cite{Long1993RankAI,Petersen1,Giov}, collaboration \cite{Giov,Wal,Lee,Glänzel}, cumulative advantages also known as the Matthew effect \cite{merton,Petersen2, Perc}, and research networks \cite{Raphael,henry2020networking}. The issue of gender in science and its correlation with success has also attracted a lot of attention \cite{henry2020networking,Kat,Yu,Long1993RankAI,Box-Steffen,Gaule,Yutao,JensP,kaminski,kwiek2024, Alessandro}, as has the rise of teams in research \cite{Wuchty,Sta,Zzeng,Pavlidis}, and other indirect factors such as the prestige of advisors, institutions, reputation of researchers, etc. \cite{su,ScottLong,DanielB,EbbeK,HugoHorta,LiW,Peters}. 

However, an important measure of academic success must be the ability of authors to maintain a long career, that is, to \textit{survive} professionally. Surprisingly few studies \cite{kaminski,kwiek2024,Freeman,Shibayama,Price,Milojevic,Boothby} have focused on this aspect, and fewer still have attempted to chart how competition and academic survival rates have changed over time \cite{Price,Milojevic,Boothby,kwiek2024}. 
That researchers `leave science' has been the subject of a number of mostly survey-based studies \cite{Aldo,ZhouY,Dorenkamp, White}, but under the assumption that this is a `voluntary act' rather than the consequence of being forced out by a competitive environment. The role of competition has also played an indirect role in a number of studies on corrupt incentive structures \cite{Stephan,Alberts} and competition between universities \cite{Marginson,Münch}. Concerning physics, a small number of bibliometrical studies \cite{Raphael,Deville,Pan,Radicchi} exist but none of which focus on competition.

That academia has become increasingly competitive is nevertheless evident as the substantial growth in the number of awarded PhD degrees \cite{Cyranoski} and researchers in temporary positions have not been matched by a similar 
expansion in the number of permanent faculty positions \cite{MW,Schille,germany}. 
This transformation may have induced altered incentive structures and selection pressures that impact the scientific community on multiple levels.   
Charting the historical changes in researcher career lengths and survival determinants is necessary for understanding when and how.

In a groundbreaking paper by Milojevic et al. \cite{Milojevic} 
the authors analyzed researcher career paths in astronomy, ecology, and robotics from 1960 to 2010 and found a dramatic shortening of the average career length in all three disciplines. While statistically significant trends between survivability and early productivity of authors were observed, neither productivity, citations of early work, or levels of collaboration emerged as reliable predictors of survivability. Despite its innovative nature this study has a few significant shortcomings. Firstly, its sampling is based solely on publication data from a list of principal journals from which the associated author metrics (lifetimes, productivity etc.) are also derived. This does not take into account the significant increase in publications, researchers, and available science journals over time and thus carries a risk of incompleteness in the metadata used for estimating the metrics. Secondly, its segmentation within the respective disciplines is non-existent, including, e.g., both theoretical and experimental researchers who may organize themselves in teams of vastly different sizes and with different research cultures and publication patterns, leading to 
risk of data inhomogeneity. 

Another study by Kwiek \& Szymula \cite{kwiek2024} investigated cohort attrition from 2000-2010 in 16 STEMM disciplines across 38 OECD countries with authors assigned to disciplines based 
on the references in their 
publications. Exponential attrition was observed for both the 2000 
and 2010 
physics cohorts with approximately 50\% of authors leaving the field after a decade. An obvious constraint of this study, however, is its limited time series, hindering the capture of long-term evolutionary changes.
%
Coarse segmentation, lack of data homogeneity, and limited time series also plague other studies \cite{kaminski, Boothby} that focus on competition and career length. 

This paper seeks to determine the historical levels of academic survival within theoretical high-energy physics (hep-th) and analyze its 
impact on the research community. For this purpose we study the careers of 30,149 researchers within hep-th from three different samples covering the period from 1950 to 2020. Using a cohort-based approach we chart the survival ratios (cohort attrition) of hep-th authors, and assess its correlation with author productivity, citations, and collaborations. A central component is our sampling method, which selects only cohorts of authors who published papers within mainstream hep-th disciplines during the first three years of their academic career. This ensures a homogeneous population based on early scientific output and allows for a more accurate assessment of the shifting trends within this population over time.

We chose hep-th as the subject of our study due to its recent historical developments. Since the discovery of the standard model of particle physics in the 1970s hep-th has been in an unprecedented situation, where it has made no major breakthroughs on the key challenges the field is faced with\footnote{For instance the problems of quantising gravity, rigorously defining non-perturbative quantum field theory, or determining the origin of the structure of the standard model of particle physics.} while at the same time has had very little empirical data to work from \cite{Ritson}. Parallel to this, the field has coalesced into a small number of large research communities centered around a set of fairly old ideas; e.g., supersymmetry \cite{Gervais:1971ji} (1971), string theory \cite{Scherk:1974ca} (1974), loop quantum gravity \cite{Ashtekar:1986yd} (1986), and AdS/CFT correspondence \cite{Maldacena:1997re} (1997) – none of which have led to any hard empirical evidence and some of which have been partly refuted experimentally \cite{Mitsou:2023dgf}. This is in stark contrast to the physics community response at the dawn of the 20th century, where discrepancies between theory and experiment led to a veritable explosion of new, radical, and ambitious ideas (e.g., quantum mechanics and general relativity). Our aim is, in part, to understand whether this contrast might stem from contemporary researchers adapting to survive in an increasingly competitive research environment.

\section{Data collection}
\subsection{Scope and sampling.}
To assess the impact of competition on mainstream research in hep-th, we extracted the full career publications of authors in various physics topics over different historical periods. The sampling was based on the following criteria:
\begin{enumerate}
    \item The topics were of a highly theoretical nature and distinguished by significant levels of research activity over the entire period.
    \item The topic was the primary research area of all authors in the sample at the beginning of their scientific career.
\end{enumerate}
Criteria 1 ensures sampling of mainstream areas of hep-th. Combined with Criteria 2 we obtain a population of hep-th authors based on their early career research output.

\subsection{Data source and processing.}
To extract the data relevant for our scope we queried OpenAlex\footnote{OpenAlex is a scientific knowledge graph launched in January 2022 to replace the discontinued Microsoft Academic Graph (MAG). See, e.g., \cite{Scheidsteger} for an overview.} for all articles published in science journals, with the filters in Table \ref{table:Samples} applied.\footnote{Note that the ``time''-filters in our initial OpenAlex search covered a slightly longer period than those specified in Table \ref{table:Samples}. During our data analysis the historical period was narrowed due to insufficient early observations for modeling purposes, see Appendix \ref{datacorr} for details. Over the course of writing this paper, the OpenAlex ``concepts'' classification system was also deprecated in favor of ``topics''. However, the topic classification system was less transparent for our segmentation purposes and we chose to continue sampling based on the concepts hierarchy.} The search field was global, i.e., we did not filter on geographical author locations. The resulting 
samples, which are denoted by $\mathcal{D}_x$, 
$x\in \{\mbox{QED},\mbox{QRSS},\mbox{STQG}\}$, 
span a period of 35 years each with the total span ranging from 1950 to 2020. A number of data corrections and exclusions (described in Appendix \ref{datacorr} and the following sections) were then performed. 

The set $\mathcal{A}_x$ of all (unique) authors 
 in $\mathcal{D}_x$ were identified via the data field \textit{authorships.author.id} in OpenAlex. We chose this as an identifier due to the disambiguity associated with using author names.
\begin{figure}[H]
    \centering
    \includegraphics[width=1\linewidth]{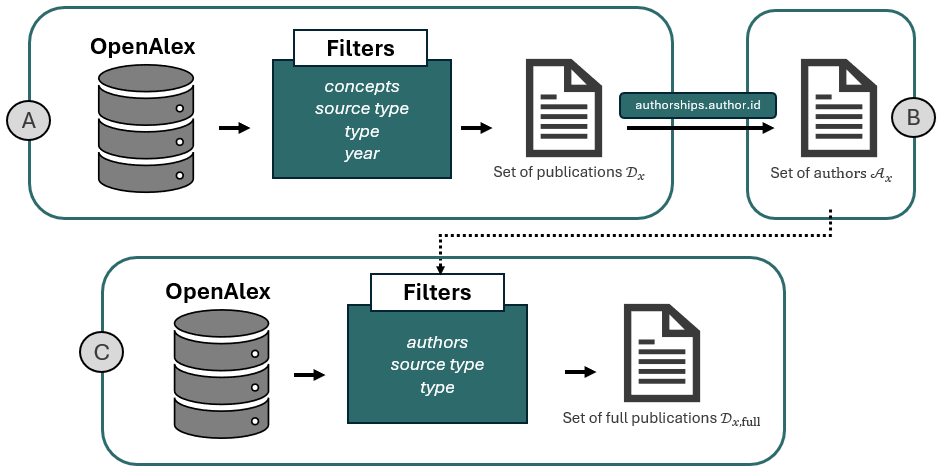}
    \caption{\small{Data collection and processing. (A) We leverage OpenAlex.org for all articles (\textit{type}) published in journals (\textit{source type}) that fall within a given physics research topic (\textit{concept}) and period (\textit{year}). This gives us a set of publication metadata $\mathcal{D}_x$. (B) From this metadata, we extract the set of all unique authors $\mathcal{A}_x$ with publications in $\mathcal{D}_x$. (C) Finally, we leverage OpenAlex again to obtain the full career publications $\mathcal{D}_{x,\text{\tiny full}}$ of the authors in $\mathcal{A}_x$. This metadata is the source from which our author metrics are derived.}}
    \label{fig:dataprocessing}
\end{figure}
\begin{table}[H]
\centering
\scalebox{0.8}{\begin{tabular}[t]{l|l|l|l}
\hline
\textbf{Author sample} & \textbf{Concept(s)} & \textbf{Year} & \textbf{N} \\
\hline
\hline
 $\mathcal{A}_{\text{\tiny QED}}$  & ``Quantum electrodynamics'' & 1950-1985 & 15,359 \\
 \hline
 $\mathcal{A}_{\text{\tiny QRSS}}$  & \begin{tabular}[x]{@{}l@{}}``Quantum field theory''\\``Renormalization group''\\``Supersymmetry''\end{tabular} & 1975-2010 & 8,570\\
\hline
$\mathcal{A}_{\text{\tiny STQG}}$   &  \begin{tabular}[x]{@{}l@{}} ``String theory''\\ ``Quantum gravity''\end{tabular} & 1985-2020 & 6,224 \\
\hline
\end{tabular}
}
\caption{\small{OpenAlex filters for generating publication samples and associated author sample sizes. For each sample the concept filters were specified by logical disjunction (OR) such that articles falling within at least one of the concepts were extracted. The ``year''-filter corresponds to the historical period. Note that the author sample sizes (N) shown here are after all data exclusions and corrections have been performed.}}
\label{table:Samples}
\end{table}
While disambiguities may still exist, we estimate the percentage of compromised authors to be small given the narrowness of our search field and extensive data corrections. All articles published by the authors in $\mathcal{A}_x$ over their entire career were then extracted from OpenAlex. We denote this set by $\mathcal{D}_{x,\text{\tiny full}}$. The overall data collection process is graphically illustrated in Fig. \ref{fig:dataprocessing}. 
 
At this point we remark that a significant proportion of authors were \textit{senior} or \textit{migrating} authors, whose first publication in the original sample $\mathcal{D}_x$ came at midpoint or late stage of their career. 
To comply with Criteria 2 of our scope these authors were excluded, as they were past the critical survival phase of their career at the time of their emergence in $\mathcal{D}_x$. This correction is described further in Appendix \ref{datacorr} and the following section. Based on the publication data in $\mathcal{D}_{x,\text{\tiny full}}$ we derived the metrics (lifetimes, productivity etc.) for the remaining authors in $\mathcal{A}_x$. The final author sample sizes after all corrections and exclusions have been applied are listed in Table \ref{table:Samples}.

\section{Methodology}
\subsection{Author career lifetimes.}
For each author $r \in \mathcal{A}_x$, we identify the first year $t_{i}(r)$ and last year $t_{f}(r)$ that the author appeared on a publication in $\mathcal{D}_{x,\text{\tiny full}}$ regardless of authorship role\footnote{Note that authors in hep-th are traditionally ordered alphabetically.} (lead or supporting). 
Since $\mathcal{D}_{x,\text{\tiny full}}$ comprises the full career publications for each author both $t_{i}(r)$ and $t_{f}(r)$ can be reliably determined. The author is then assigned a \textit{total career lifetime} of 
\begin{equation*}
    T(r) := t_{f}(r) - t_{i}(r) +1, 
\end{equation*}
where 1 is added to ensure a minimum lifetime of one year. All authors in $\mathcal{A}_x$ are then grouped into cohorts $\mathcal{C}_{x,y}$ based on their first year of publication:
\begin{equation*}
    \mathcal{C}_{x,y} := \lbrace r \in \mathcal{A}_x \ | \ t_i (r) = y \rbrace.
\end{equation*}
To comply with Criteria 2 in our scope, however, we consider only the subset $\mathcal{C}_{x,y}^{\text{\tiny{sub}}}$ of authors in each cohort whose first publication in $\mathcal{D}_x$ came within the first $\Delta$ years of their academic career. Formally:
\begin{equation*}
    \mathcal{C}_{x,y}^{\text{\tiny{sub}}} := \lbrace r \in \mathcal{C}_{x,y} \ | \ \widehat{t}_{i}(r) - t_{i}(r) \leq \Delta \rbrace,
\end{equation*}
where $\widehat{t}_{i}(r)$  denotes the first year that the author appeared in $\mathcal{D}_x$. We shall call this a \textit{cohort} for the remainder of the paper (and drop the subscript $x$ for brevity). For this study we set $\Delta =3$ as this represents the typical duration of a PhD where junior researchers specialize in a topic. Finally, we identify and exclude \textit{transient} authors \cite{Price}, i.e., authors where $T(r) = 1$. This is in line with \cite{Milojevic}. The final cohort sizes $\mathcal{C}_{y}^{\textnormal{\tiny{sub}}}$ for each sample (QED, QRSS, STQG) are shown in Appendix \ref{cohortsizes}.

We note an overlap 
between the samples.
\footnote{Approximately 20-25\% of the authors in the QRSS cohorts are present in the STQG cohorts throughout, except for a noticeable drop from 2008-2010. Approximately 8-16\% of the authors in the QED cohorts are also present in the QRSS cohorts during the overlap period from 1975-1985.} This is unavoidable within our scope and sampling method as, e.g., supersymmetry plays a major role in string theory. We made no adjustments to account for this as any exclusion of mutually shared authors between the samples would induce a bias.
\subsection{Survival rates.} To model changes in the academic survival rates over time, we compute the cumulative distribution $|E_y (\Lambda)|$ of authors in a cohort $\mathcal{C}_y^{\text{\tiny{sub}}}$ whose total lifetime does not exceed $\Lambda \in \mathbb{N}$, i.e.,
\begin{equation*}
    E_y (\Lambda) = \lbrace r \in \mathcal{C}_y^{\text{\tiny{sub}}} \ | \ T(r) \leq \Lambda \rbrace,
\end{equation*}
where $|E_y (\Lambda)|$ denotes the size of the set. The $\Lambda$-year \textit{survival function} $S_y (\Lambda)$ is then
\begin{equation*}
    S_y (\Lambda) = 1-\frac{|E_y (\Lambda)|}{|\mathcal{C}_y^{\mbox{\tiny sub}}|},
\end{equation*}
and represents the fraction of the cohort $y$ remaining after $\Lambda$ years.\footnote{Given that we are working with right-censored data, we note that the survival analysis is biased near the tail-end of the time series for the most recent sample $\mathcal{D}_{\text{\tiny{STQG}}}$ as authors whose last publication falls a few years short of 2020 need not necessarily have perished, but may simply be inactive for the moment.} We additionally compute the survival function for any given cohort using the Kaplan-Meier estimator. This is a nonparametric statistic often used to model survival distribution functions in censored data. For any cohort $\mathcal{C}_y^{\mbox{\tiny sub}}$ the Kaplan-Meier estimator $\text{KM}_y (t)$ is
\begin{equation*}
    \text{KM}_y (t) = \prod_{t_j \leq t} \left(1 -\lambda(t_j) \right).
\end{equation*}
Here, $\lambda(t_j)$ are the discrete-time hazard rates measuring the probability of surviving beyond time $t_j$ conditioned on having survived up to time $t_j$.

%
%

\subsection{Author classification.} Based on the empirical data, we construct a binary classifier
\begin{equation} \label{survivalfunction}
\text{SURV}(r) :=
\begin{cases}
1 & \text{if } T(r) > T_{\text{\tiny{comp}}},\\
0 & \text{if } T(r) \leq T_{\text{\tiny{comp}}},
\end{cases}
\end{equation}
where $T_{\text{\tiny{comp}}}$ is a survival threshold, and group authors in each cohort $\mathcal{C}_y^{\mbox{\tiny sub}}$ as either \textit{surviving} or \textit{non-surviving}. Naturally, this method poses a challenge in assessing the appropriate choice of threshold $T_{\text{\tiny{comp}}}$. Ideally, this should reflect the period until researchers acquire permanent faculty position or the point at which academic competition becomes insignificant. Concerning the former, little public information is available about the mean duration that researchers spend in temporary positions. A 2018 workforce survey from the AAS (American Astronomical Society) estimates that the average cumulative postdoctoral period among tenured members was 4 years \cite{workforce}. An analysis conducted by the University of Toronto, based on data from the Theoretical Particle Jobs Rumor Mill, indicates that the median post-doc duration in hep-th was 4-6 years from 1994-2017 \cite{toronto}.  Additionally, data from the German Statistisches Bundesamt (Destatis) show that the mean age of researchers obtaining habilitation in physics or astronomy in Germany was around 40 years from 1992-2021 \cite{germany}. Based on these observations, we fix $T_{\text{\tiny{comp}}}=12$ in the following as we judged this to be an accurate representative of the duration where most researchers are actively competing for survival. Naturally, however, the true value of $T_{\text{\tiny{comp}}}$ is likely to evolve over time, by location, or even at the level of individual institutions. We postpone attempts to estimate a dynamical value, but remark that changing $T_{\text{\tiny{comp}}}$ to $8$ or $10$ years has insignificant impact on the general trends observed in this paper.
\subsection{Early indicators of author survivability.} \label{desstats}
We investigate early career characteristics of researchers classified according to Eq. \eqref{survivalfunction}. We focused on variables directly related to the production of scientific knowledge: Author productivity, impact, and collaboration. These variables have been found to be strongly correlated with researcher career trajectories in numerous other studies \cite{Sinatra2,LiuL,Xiaomei, Peters,Long1993RankAI,Petersen1,Giov,Wal,Lee,Glänzel}. Specifically, we investigate the early productivity rate $\text{prod}_{{\text{\tiny{rate}}}} (r)$, early collaborator rate $\text{col}_{{\text{\tiny{rate}}}} (r)$ and early citation rate $\text{cit}_{{\text{\tiny{rate}}}} (r)$, defined for each author
$r$ as
\begin{align*} 
    &\text{prod}_{{\text{\tiny{rate}}}} (r) := \frac{P_{\text{\tiny{early}}}(r)}{T},  \\
    &\text{col}_{{\text{\tiny{rate}}}} (r) := \frac{\text{col}_{{\text{\tiny{early}}}}(r)}{T},  
    \\ &\text{cit}_{{\text{\tiny{rate}}}} (r)  :=  \frac{\text{cit}_{{\text{\tiny{early}}}}(r)}{T}, 
\end{align*}
where 
\begin{align*}
    T:=\begin{cases}
    T(r) & \text{if } T(r) \leq T_{\text{\tiny{comp}}},\\
    T_{\text{\tiny{comp}}} & \text{if } T(r) > T_{\text{\tiny{comp}}}.
    \end{cases}
\end{align*}
Here, $P_{\text{\tiny{early}}}(r)$ is the total number of papers published by an author $r$ until time $T$ measured using the full count method, where each coauthor of a paper is assigned a credit of one. Meanwhile, $\text{col}_{{\text{\tiny{early}}}}(r)$ is the total number of (non-unique) coauthors that an author $r$ has up to time $T$, and $\text{cit}_{{\text{\tiny{early}}}} (r)$ is the total number of citations generated by papers coauthored by an author $r$ up to time $T$. 

Additionally, to avoid collaboration bias, we examined the early fractional productivity rate $\text{fprod}_{{\text{\tiny{rate}}}} (r)$ and early citations per paper $\text{C}_{\text{p}}\text{P} (r)$ defined as
\begin{subequations}
    \begin{align*}
    &\text{fprod}_{{\text{\tiny{rate}}}} (r) := \frac{P_{\text{\tiny{early, frac}}}(r)}{T},  
    \\
    &\text{C}_{\text{p}}\text{P} (r) :=  \frac{\text{cit}_{{\text{\tiny{early}}}} (r)}{P_{{\text{\tiny{early}}}} (r)},  
\end{align*}
\end{subequations}
where $P_{\text{\tiny{early, frac}}}$ is the total number of publications by author $r$ until time $T$ measured using an unweighted fractional count, where each coauthor of a paper is credited with a count of $\frac{1}{\text{total number of authors}}$. 

The use of \textit{rates} (rather than absolute counts) was chosen in order to normalize the effects of varying lifetimes among short-lived authors, given that the total early collaborators, papers authored, and citations, is likely to increase with career length.  

Shifts in the variable distributions over time were estimated using the Population Stability Index (PSI). This is a modified version of the Kullback-Leibler divergence and a widely used measure for assessing distributional shifts between binned samples. Given a reference distribution $A$ and a target distribution $B$ both are partitioned into $K$ bins based on the percentiles of $A$. The PSI is then given by the formula
\begin{equation*}
    \text{PSI}(A,B)= \sum_{i=1}^K (\%A_i - \%B_i) \ln \left(\frac{\% A_i}{\% B_i} \right),
\end{equation*}
where $\%A_i$ and $\%B_i$ are the proportions of observations in bins $A_i$ and $B_i$, $i\in\{1,\ldots, K\}$, respectively. As common thresholds $\text{PSI}(A,B)<0.1$ indicates similarity of the distributions, $0.1 \leq \text{PSI}(A,B)<0.25$ indicates a moderate difference, and $\text{PSI}(A,B) \geq 0.25$ a significant difference in terms of skewness or asymmetry \cite{Siddiqi}.
\subsection{Predictive model.} \label{prediction}
To model author survivability, a logistic regression was fitted to the survival data in our samples using the SURV-function in Eq. \ref{survivalfunction} with $T_{\text{\tiny{comp}}}=12$ as the response variable:
\begin{equation*}
    \mathbb{P}(\text{SURV}=0) = \frac{1}{1+e^{-(\beta_0 + \sum_{j=1}^n \beta_j x_j)}}.
\end{equation*}
Here, $x_j$ are the predictor variables, $\beta_0$ the intercept and $\beta_j$ the parameters of the function, respectively. The latter were obtained by Maximum Likelihood Estimation (MLE), which finds the parameter values maximizing the likelihood of observing the given outcome. 

Predictor variables were extracted by a stepwise forward selection procedure beginning with a univariate logistic regression where all key variables described in section \ref{desstats} were tested. To ensure robust estimates, a bootstrap method was employed, whereby the data samples were randomly partitioned into a development sample and a testing sample based on an 80/20-split (k-fold cross validation was disregarded due to data limitations). The model was then fitted and tested over 10,000 iterations. Univariate entry selection criteria were based on a combination of variable significance ($p<0.05$) and discriminatory power. The latter was estimated by the area Under the Receiver Operating Characteristics (AUROC) curve. Only univariate models with $\text{AUROC}>0.65$ were considered for further analysis. Subsequently, multivariate analysis was conducted with retention criteria based on variable significance ($p<0.05$), direction and magnitude of the parameter values, and discriminatory power. Additionally, covariance of the predictors was assessed and multicollinearity checked via the Variance Inflation Factor (VIF). We did not investigate models with both $\text{prod}_{\text{\tiny{rate}}}$ and $\text{fprod}_{\text{\tiny{rate}}}$, or $\text{cit}_{\text{\tiny{rate}}}$ and $\textnormal{CpP}$ as predictors, due to their similarity. The process including grouping of data for modeling purposes is described further in section \ref{modellingsec}.

\section{Results}  

\subsection{Survival rates.}

Survival curves for researcher cohorts are shown in

\begin{figure}[H]
    \centering
    \includegraphics[width=1\linewidth]{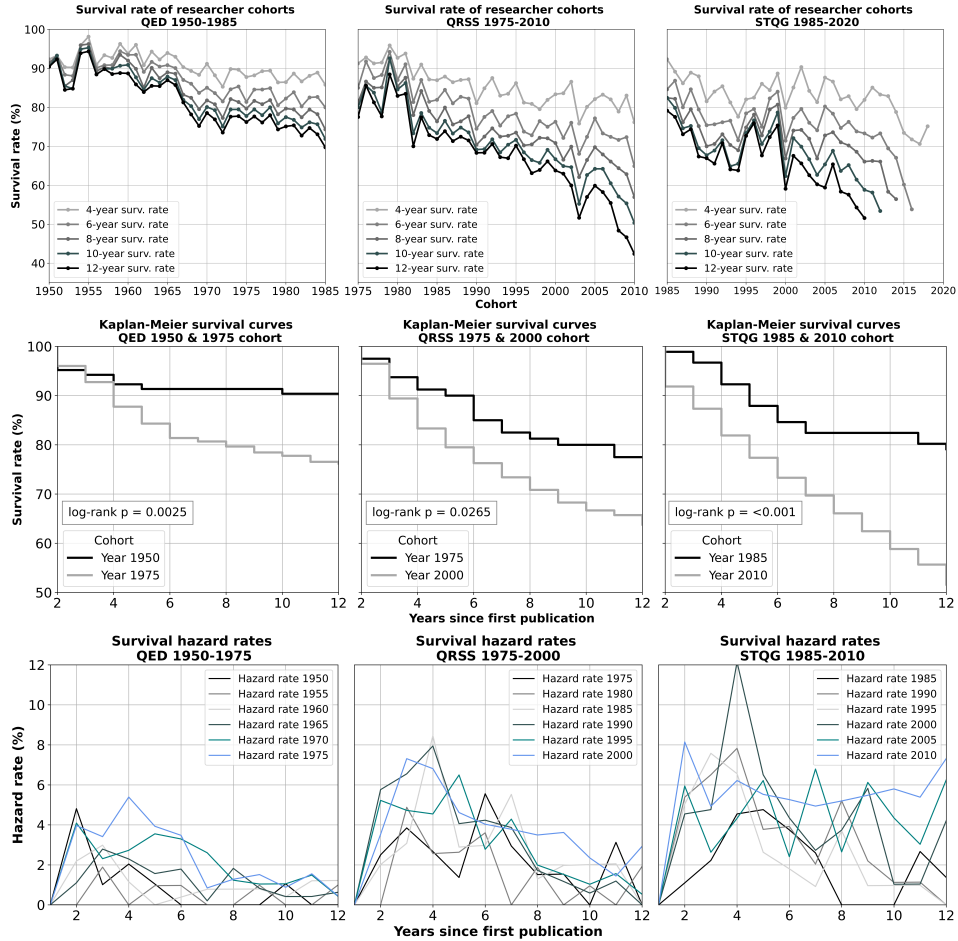}
    \caption{\small{Survival rates, hazard rates, and Kaplan-Meier survival curves. Top: Survival rates of cohorts in the QED sample 1950-1985 (left), the QRSS sample 1975-2010 (middle) and the STQG sample 1985-2020 (right). Survival curves for the latter STQG cohorts are truncated due to right-censoring of the data. Middle: Kaplan-Meier curves measured over 12 years for the first cohort in the respective samples, compared to the cohort 25 years later. Results of the log-rank test are shown. Bottom: Hazard rates for different cohorts separated by 5-year intervals and measured over a 12-year period for each sample. The vanishing hazard rate at year one stems from the fact that all authors with a one-year lifetime have been removed.}}
    \label{fig:surv_merged}
\end{figure}

\noindent Fig. \ref{fig:surv_merged} (top) for all three samples. We observe an evolution from relatively high and steady survival rates prior to the 1980s, toward a significant downtrend accelerating from the 2000s onward. 

While the 12-year survival rate remains above 70\% for all QED and QRSS cohorts prior to 1980, even lingering above 90\% in the 1950s-1960s for QED, levels below 70\% are seen several times during the 1990s and below 50\% for the QRSS cohorts in the 2000s. The results indicate a rapid, continuous decline in the expected career lengths of researchers over the past four decades. We also observe an increasing divergence in the survival curves over time: 
While the 4-, 6-, 8- and 10-year curves generally follow the same trend as the 12-year curve, their slopes differ, most noticeably beginning in the 1990s (QRRS) and 2000s (STQG) for the 4- and 6-year curves, indicating significant researcher dropout around the years after PhD and early postdoc studies are expected to have been completed. 

In Fig. \ref{fig:surv_merged} (middle), the Kaplan-Meier survival curves for the earliest cohort in the samples and the cohort 25 years later are compared. Curves for the latter cohorts are marked by a steady year-after-year population decline, while curves for the early cohorts appear to stabilize around an almost constant population after a number of years (approximately 4 years for the QED 1950 cohort and 7-8 years for the QRSS 1975 and STQG 1985 cohort). Additionally, the log-rank test rejects the null hypothesis of similar survival distributions between the first and last cohorts in the 25-year span at 95\%-confidence level ($p<0.05$). Investigating the hazard rates, shown in Fig. \ref{fig:surv_merged} (bottom), we likewise observe stabilization below 2\% after 4-8 years for all 1950-1975 cohorts within the QED sample, with the hazard rate never going above 6\%. Similar patterns are observed for the QRSS cohorts after 7-10 years from 1975-2000, but with hazard rates reaching up to 8\% early on. For the STQG sample, higher hazard rates are observed for all cohorts, but with stabilization after 8-10 years prior to the 2000s. For the 2000s onward, no tendency to stabilize around a constant rate is observed, with peak levels of 12\% for the STQG 2000 cohort and rates above 4\% for the 2005 and 2010 cohort even after 8-12 years.

We remark that the survival rates obtained in this section are upper bounds. Due to our sampling method we can only distinguish survival with respect to remaining in academia at large, not hep-th specifically. 
\begin{figure}[H] 
    \centering
    \includegraphics[width=1\linewidth]{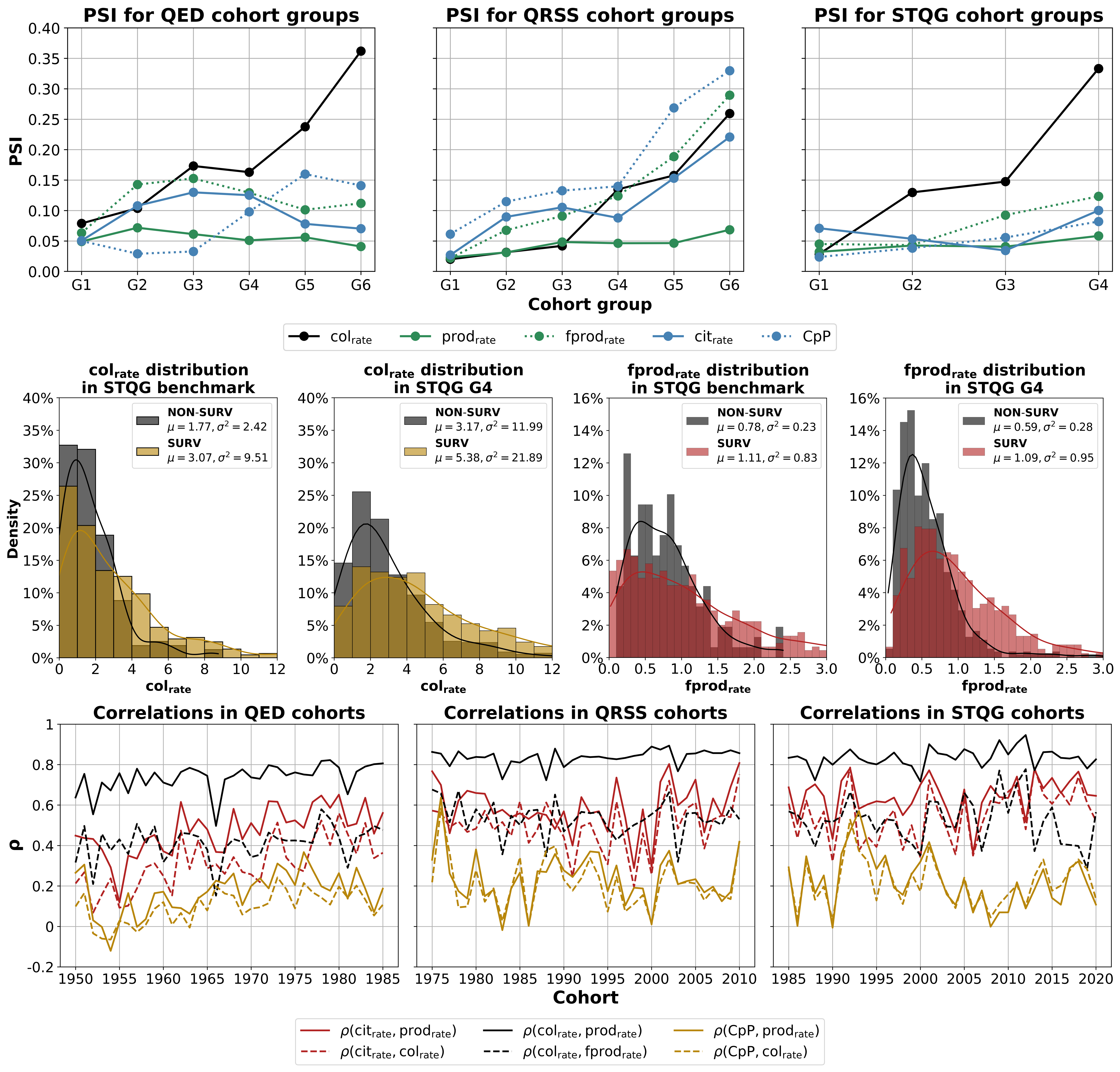}
    \caption{\small{Distributional shifts and covariance of key variables for cohort groups. Top: PSI levels for cohort groups, indicating distributional shifts of key variables in each sample over time. Middle: PMF (probability mass function) with KDE (kernel density estimation) for $\text{col}_{{\text{\tiny{rate}}}}$ and $\text{fprod}_{{\text{\tiny{rate}}}}$ among survivors and non-survivors in the STQG benchmark (1985-1989) and G4 (2005-2010) cohort groups. Bottom: Pearson correlation coefficients for different combinations of $\text{col}_{{\text{\tiny{rate}}}}$, $\text{prod}_{{\text{\tiny{rate}}}}$, $\text{fprod}_{{\text{\tiny{rate}}}}$ $\text{cit}_{{\text{\tiny{rate}}}}$, and $\text{C}_{\tiny{\text{p}}}\text{P}$ at the cohort level for each sample.}}
    \label{fig:PSIcorr}
\end{figure}
\subsection{Descriptive statistics of key variables.} 

Covariance of key variables over time is assessed by computing the Pearson correlation coefficient
\begin{equation*}
    \rho (x,y) = \frac{\sum_{r=1}^n (x_r - \overline{x})(y_r - \overline{y})}{\sqrt{\sum_{r=1}^n (x_r - \overline{x})^2}\sqrt{\sum_{r=1}^n (y_r - \overline{y})^2}}
\end{equation*}
for variables $x$ and $y$ at the cohort level. Here, $n$ denotes the cohort size. The results are shown in Fig. \ref{fig:PSIcorr} (bottom). Significant positive correlation between $\text{col}_{\text{\tiny{rate}}}$ and $\text{prod}_{\text{\tiny{rate}}}$ is observed with levels above 0.8 for most cohorts in the QRSS and STQG samples. Positive covariance is also consistent for the other variables, but with more moderate levels and noisier throughout. The results indicate potential collinearity of $\text{col}_{{\text{\tiny{rate}}}}$ and $\text{prod}_{{\text{\tiny{rate}}}}$. This is not surprising, given that the number of early publications by an author (measured using the full count method) is likely to increase through collaborative efforts. 

To ensure sufficient observations for the PSI test, cohorts in each sample are pooled into groups, shown in Table \ref{table:binning}. 
\begin{table}[H]
\centering
\scalebox{0.8}{
\begin{tabular}[t]{c|lll}
\hline
\textbf{Group name} & \textbf{QED cohorts} & \textbf{QRSS cohorts} & \textbf{STQG cohorts} \\
\hline
\hline
& 1950-1954 & 1975-1979 & 1985-1989 \\
Benchmark & $N_{\text{\tiny{surv}}}=447$ & $N_{\text{\tiny{surv}}} =390$ & $N_{\text{\tiny{surv}}} =449$ \\
& $N_{\text{\tiny{non-surv}}}=54$ & $N_{\text{\tiny{non-surv}}} =82$ & $N_{\text{\tiny{non-surv}}} =159$ \\
\hline
& 1955-1959 & 1980-1984 & 1990-1994 \\
G1 & $N_{\text{\tiny{surv}}}=766$ & $N_{\text{\tiny{surv}}} =618$ & $N_{\text{\tiny{surv}}} =413$ \\
& $N_{\text{\tiny{non-surv}}}=89$ & $N_{\text{\tiny{non-surv}}} =187$ & $N_{\text{\tiny{non-surv}}} =210$ \\
\hline
& 1960-1964 & 1985-1989 & 1995-1999 \\
G2 & $N_{\text{\tiny{surv}}}=1,536$ & $N_{\text{\tiny{surv}}} =879$ & $N_{\text{\tiny{surv}}} =512$ \\
& $N_{\text{\tiny{non-surv}}}=256$ & $N_{\text{\tiny{non-surv}}} =338$ & $N_{\text{\tiny{non-surv}}} =191$ \\
\hline
& 1965-1969 & 1990-1994 & 2000-2004 \\
G3 & $N_{\text{\tiny{surv}}}=2,387$ & $N_{\text{\tiny{surv}}} =830$ & $N_{\text{\tiny{surv}}} =524$ \\
& $N_{\text{\tiny{non-surv}}}=552$ & $N_{\text{\tiny{non-surv}}} =386$ & $N_{\text{\tiny{non-surv}}} =307$ \\
\hline
& 1970-1974 & 1995-1999 & 2005-2010 \\
G4 & $N_{\text{\tiny{surv}}}=2,300$ & $N_{\text{\tiny{surv}}} =974$ & $N_{\text{\tiny{surv}}} =756$ \\
& $N_{\text{\tiny{non-surv}}}=695$ & $N_{\text{\tiny{non-surv}}} 505$ & $N_{\text{\tiny{non-surv}}} =551$ \\
\hline
& 1975-1979 & 2000-2004 & \\
G5 & $N_{\text{\tiny{surv}}}=2,037$ & $N_{\text{\tiny{surv}}} =961$ & - \\
& $N_{\text{\tiny{non-surv}}}=628$ & $N_{\text{\tiny{non-surv}}} =670$ & \\
\hline
& 1980-1985 & 2005-2010 & \\
G6 & $N_{\text{\tiny{surv}}}=2,647$ & $N_{\text{\tiny{surv}}} =933$ & - \\
& $N_{\text{\tiny{non-surv}}}=962$ & $N_{\text{\tiny{non-surv}}} =817$ & \\
\hline
\end{tabular}
}
\caption{\small{Grouping of cohorts. To calculate the PSI statistic, cohorts are pooled together in groups of five-year intervals (six for the final group), resulting in a total of seven groups for the QED and QRSS samples, and five groups for the STQG sample. In each case the benchmark group for the PSI statistic is the group containing the earliest five cohorts in the sample. The total number of survivors ($N_{\text{\tiny{surv}}}$) and non-survivors ($N_{\text{\tiny{non-surv}}}$) in each group is shown.}}
\label{table:binning}
\end{table}
Each group is then partitioned into bins based on the 10th percentile 
\begin{figure}[H] 
    \centering
    \includegraphics[width=1\linewidth]{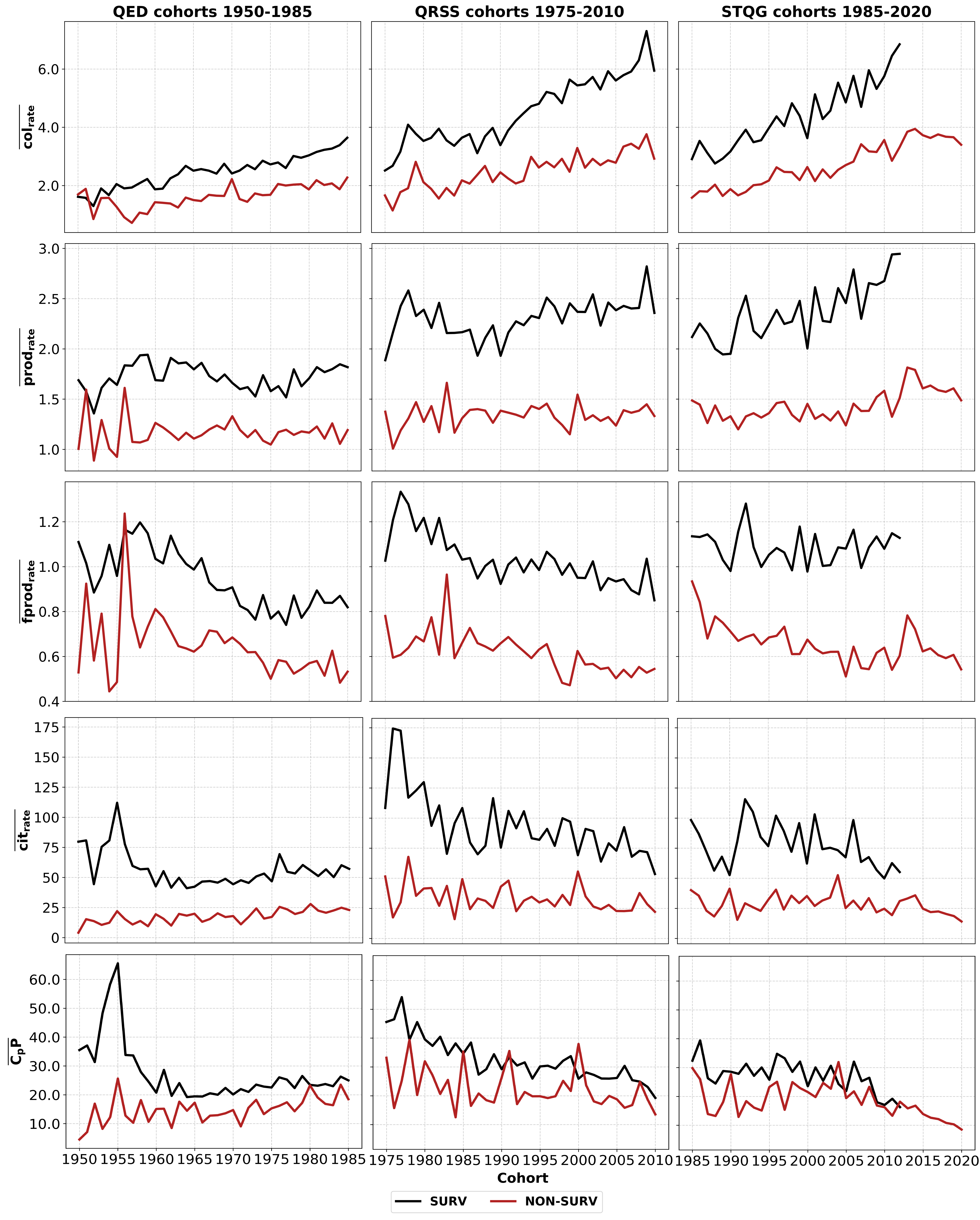}
    \caption{\small{Evolution of variable means over time. Evolution of the mean early collaboration rates $\overline{\text{col}_{\text{\tiny{rate}}}}$, early productivity rates (full count $\overline{\text{prod}_{\text{\tiny{rate}}}}$ and fractional count $\overline{\text{fprod}_{\text{\tiny{rate}}}}$), early citation rates $\overline{\text{cit}_{\text{\tiny{rate}}}}$ and citations per early paper $\overline{\text{CpP}}$ among surviving and non-surviving authors in each cohort.}}
    \label{fig:meansmerged}
\end{figure}
%
%
\noindent
of the variable distributions in the benchmark group, which in each case is the group consisting of the earliest five cohorts in the sample. The PSI levels are shown in Fig. \ref{fig:PSIcorr} (top). The most significant shift is observed for $\text{col}_{\text{\tiny{rate}}}$, with almost monotonically increasing PSI levels over time and final levels above 0.3 for all three samples. Remarkably, $\text{prod}_{\text{\tiny{rate}}}$ appears relatively stable, with no PSI levels above 0.1, indicating no distributional shift in early author productivity rates in hep-th since the 1950s. Contrary, moderate to significant shifts are observed for $\text{fprod}_{\text{\tiny{rate}}}$ across all samples. We note that all variables except $\text{prod}_{\text{\tiny{rate}}}$ in the QRSS sample exhibit moderate to significant shifts.

Investigating the distributions among survivors and non-survivors, we note that the mean and dispersion of $\text{col}_{\text{\tiny{rate}}}$ has increased for both in all three samples, resulting in an overall right-shift and wider distribution over time. While the mean and variance of $\text{fprod}_{\text{\tiny{rate}}}$ has not changed significantly, the distribution has become more symmetrical and approached a log-normal shape. As the number of survivors and non-survivors in each sample has generally increased over time, this may be partially attributed to the Law of Large Numbers. We note, however, that the tendency is also apparent in the STQG sample, where cohort sizes do not increase significantly over time (see Appendix \ref{cohortsizes}). Probability density plots for $\text{col}_{\text{\tiny{rate}}}$ and  $\text{fprod}_{\text{\tiny{rate}}}$ in the benchmark (1985-1989) and G4 (2005-2010) STQG groups are shown in \ref{fig:PSIcorr} (middle) for comparison. Density plots for the benchmark and the newest cohort groups in the other samples are shown in Appendix \ref{vardistributions}.

Turning our attention to differences between surviving and non-surviving authors in each cohort, we compute the mean of the key variables for the surviving and non-surviving author groups, respectively. The resulting times series are shown in Fig. \ref{fig:meansmerged}. We note an increasing trend in $\overline{\text{col}_{{\text{\tiny{rate}}}}}$ over time for both groups, with a marked discrepancy from the early 1990s onward, where levels among survivors are significantly higher. Overall, $\overline{\text{col}_{{\text{\tiny{rate}}}}}$ has roughly doubled for non-surviving authors in hep-th and increased more than threefold for surviving authors from the 1950s until the 2010s. Levels for $\overline{\text{prod}_{{\text{\tiny{rate}}}}}$ and $\overline{\text{cit}_{{\text{\tiny{rate}}}}}$ are consistently higher among survivors, the former with an increasing, and the latter with a decreasing trend across the samples, likely due to the effect of citations accumulating over time. For $\overline{\text{CpP}}$, discrepancies between the groups almost vanish and we likewise observe an overall decreasing trend. For $\overline{\text{fprod}_{{\text{\tiny{rate}}}}}$, discrepancies between the groups persist over time, but with no clear trend. 
\subsection{Modeling author survivability.} \label{modellingsec}
Modeling author survivability by logistic regression is naturally challenged by the volume and grouping of our data. At the level of individual cohorts we do not have sufficient observations to ensure robust modeling. Conversely, modeling at the sample level means fitting on data spanning 35 years with considerable variable shifts, while also preventing us from studying the evolution of predictor strength over time. To achieve both requisites we pooled cohorts into the same groups used for the PSI test (see Table \ref{table:binning}) excluding groups with less than 1,000 observations. A logistic regression was then fitted to the data in the QRSS G6 (2005-2010) and STQG G4 (2005-2010) separately. Results of the univariate analysis for all key variables are provided in Appendix \ref{univariateanalysis}. 

The final two-factor model from our step-forward procedure (see Section \ref{prediction}) using $\text{col}_{\text{\tiny{rate}}}$ and $\text{fprod}_{\text{\tiny{rate}}}$ as predictors was selected based on average performance on both cohort groups. The mean performance metrics over the 10,000 iterations for this model are shown in Appendix \ref{multivariateanalysis}, alongside performance of other candidate models. 

This two-factor model was then fitted on the other cohort groups in Table \ref{table:binning}. The mean results of the multivariate analysis over 10,000 iterations are shown in Table \ref{table:multivar}. With the exception of the QRSS 1990-1994 group (G4), $\text{fprod}_{\text{\tiny{rate}}}$ is found to be the more significant variable, but with greater uncertainty in the parameter estimates. Moderate positive correlation between $\text{col}_{\text{\tiny{rate}}}$ and $\text{fprod}_{\text{\tiny{rate}}}$ is observed across all groups, but in each case with $\text{VIF}<5$, indicating no collinearity. Performance metrics are shown in Table \ref{table:mperformance}. We note that the ability of the model to accurately distinguish between survivors and non-survivors deteriorates going into the past, with a significant decline around the early 1990s. From the mid-1990s onward, however, discriminatory power and accuracy is moderate, both on the QRSS and STQG cohort groups. Meanwhile performance is on par with a random classifier for all QED groups despite the larger number of observations in this sample and a ratio of non-survivors above 20\%. Note that for brevity only results for the QED groups G5 (1980-1984) and G6 (1985-1989) are shown in Tables \ref{table:multivar} and \ref{table:mperformance}. 
\begin{table}[H]
\centering
\scalebox{0.7}{
\begin{tabular}[t]{l|lcccccc}
\hline
\textbf{Cohort group} & Predictor & Coef. & Odds ratio & Std. Err. & $z$ & $p>|z|$ & 95\% conf. interval \\
\hline
\hline
\textbf{QED G5}& $\text{col}_{\text{\tiny{rate}}}$& -0.08514 & 0.91848 & 0.02754 & -3.08188 &  $<0.001$ & $[-0.13911, -0.03116]$ \\
\textbf{(1975-1979)}&$\text{fprod}_{\text{\tiny{rate}}}$& -0.59570 & 0.55198 & 0.11826 & -5.03106 & $0.0046$ & $[-0.82754, -0.36386]$ \\
\hline
\textbf{QED G6}& $\text{col}_{\text{\tiny{rate}}}$& -0.11861 & 0.88822 & 0.02212 & -5.35499 &  $<0.001$ & $[-0.16196, -0.07523]$ \\
\textbf{(1980-1985)}&$\text{fprod}_{\text{\tiny{rate}}}$& -0.64445 & 0.52574 & 0.09860 & -6.52902 & $<0.001$ & $[-0.83771, -0.45120]$ \\
\hline
\textbf{QRSS G2}& $\text{col}_{\text{\tiny{rate}}}$& -0.10538 & 0.90017 & 0.03586 & -2.92415 & 0.0072 & $[-0.17565, -0.03510]$ \\
\textbf{(1985-1989)}&$\text{fprod}_{\text{\tiny{rate}}}$& -0.59609 & 0.55223 & 0.14838 & -4.01193 & $<0.001$ & $[-0.88691, -0.30526]$ \\
\hline
\textbf{QRSS G3}& $\text{col}_{\text{\tiny{rate}}}$& -0.13287 & 0.87569 & 0.03260 & -4.06742 & $<0.001$ & $[-0.19676, -0.06898]$ \\
\textbf{(1990-1994)}&$\text{fprod}_{\text{\tiny{rate}}}$& -0.58223 & 0.55982 & 0.14912 & -3.89769 & $<0.001$ & $[-0.87641, -0.28982]$ \\
\hline
\textbf{QRSS G4}& $\text{col}_{\text{\tiny{rate}}}$& -0.12020 & 0.88685 & 0.02686 & -4.46537 & $<0.001$ & $[-0.17284, -0.06758]$ \\
\textbf{(1995-1999)}&$\text{fprod}_{\text{\tiny{rate}}}$& -1.09897 & 0.33558 & 0.16229 & -6.75694 & $<0.001$ & $[-1.41705, -0.78089]$ \\
\hline
\textbf{QRSS G5}& $\text{col}_{\text{\tiny{rate}}}$& -0.14168 & 0.86799 & 0.02378 & -5.94704 & $<0.001$ & $[-0.18828, -0.09506]$ \\
\textbf{(2000-2004)}&$\text{fprod}_{\text{\tiny{rate}}}$& -0.89363 & 0.41004 & 0.14485 & -6.16488 & $<0.001$ & $[-1.17754, -0.60972]$ \\
\hline
\textbf{QRSS G6}&$\text{col}_{\text{\tiny{rate}}}$& -0.10799 & 0.89776 & 0.02037 & -5.28358 & $<0.001$ & $[-0.14791, -0.06806]$ \\
\textbf{(2005-2010)}&$\text{fprod}_{\text{\tiny{rate}}}$& -1.13116 & 0.27038 & 0.15983 & -8.20082 & $<0.001$ & $[-1.62483, -0.99831]$ \\
\hline
\textbf{STQG G4}&$\text{col}_{\text{\tiny{rate}}}$& -0.08561 & 0.91806 & 0.02550 & -3.3448 & 0.0026 & $[-0.13558, -0.03563]$ \\
\textbf{(2005-2010)}&$\text{fprod}_{\text{\tiny{rate}}}$& -1.23308 & 0.29298 & 0.16795 & -7.33141 & $<0.001$ & $[-1.56225, -0.90391]$ \\
\hline
\end{tabular}
}
\caption{\small{Multivariate analysis. Mean results of multivariate logistic regression with $\text{col}_{\text{\tiny{rate}}}$ and $\text{fprod}_{\text{\tiny{rate}}}$ as predictor variables, fitted on cohort groups with more than 1,000 data points over 10,000 iterations.}}
\label{table:multivar}
\end{table}

\begin{table}[H]
\centering
\scalebox{0.8}{\begin{tabular}[t]{l|ccccclc}
\hline
\textbf{Cohort group} & N & Youden's J & AUROC & Acc. & F1 & LR $\chi^2$ & $p>\chi^2$  \\
\hline
\hline
QED G5 (1975-1979) & 2,665 & 0.22 & 0.59 & 0.51 & 0.44 & 92.6 &$<0.001$  \\
QED G6 (1980-1985) & 3,609 & 0.26 & 0.62 & 0.56 & 0.48 & 202.8 & $<0.001$ \\
QRSS G2 (1985-1989) & 1,217 & 0.28 & 0.62 & 0.57 & 0.50 & 73.9 & $<0.001$ \\
QRSS G3 (1990-1994) & 1,216 & 0.28 & 0.63 & 0.59 & 0.55 & 97.0 &$<0.001$  \\
QRSS G4 (1995-1999) & 1,479 & 0.39 & 0.71 & 0.64 & 0.62 & 209.8 &$<0.001$  \\
QRSS G5 (2000-2004) & 1,631 & 0.36 & 0.69 & 0.65 & 0.67 & 226.5 &$<0.001$  \\
QRSS G6 (2005-2010) & 1,750 & 0.36 & 0.73 & 0.67 & 0.70 & 288.0 & $<0.001$ \\
STQG G6 (2005-2010) & 1,307 & 0.37 & 0.72 & 0.66 & 0.67 & 193.0 & $<0.001$ \\
\hline
\end{tabular}
}
\caption{\small{Performance measures for the survival model. Performance of the two-factor logistic regression model with $\textnormal{fprod}_{\textnormal{\tiny{rate}}}$ and $\textnormal{col}_{\textnormal{\tiny{rate}}}$ as predictors, tested on PSI cohort groups with more than 1,000 observations across the various samples. Note the deterioration in discriminatory power and calibration when fitted on older cohort groups. For all cohort groups in the QED sample, performance was comparable to random classifier.}}
\label{table:mperformance}
\end{table}

\begin{figure}[H] 
    \centering
    \includegraphics[width=1\linewidth]{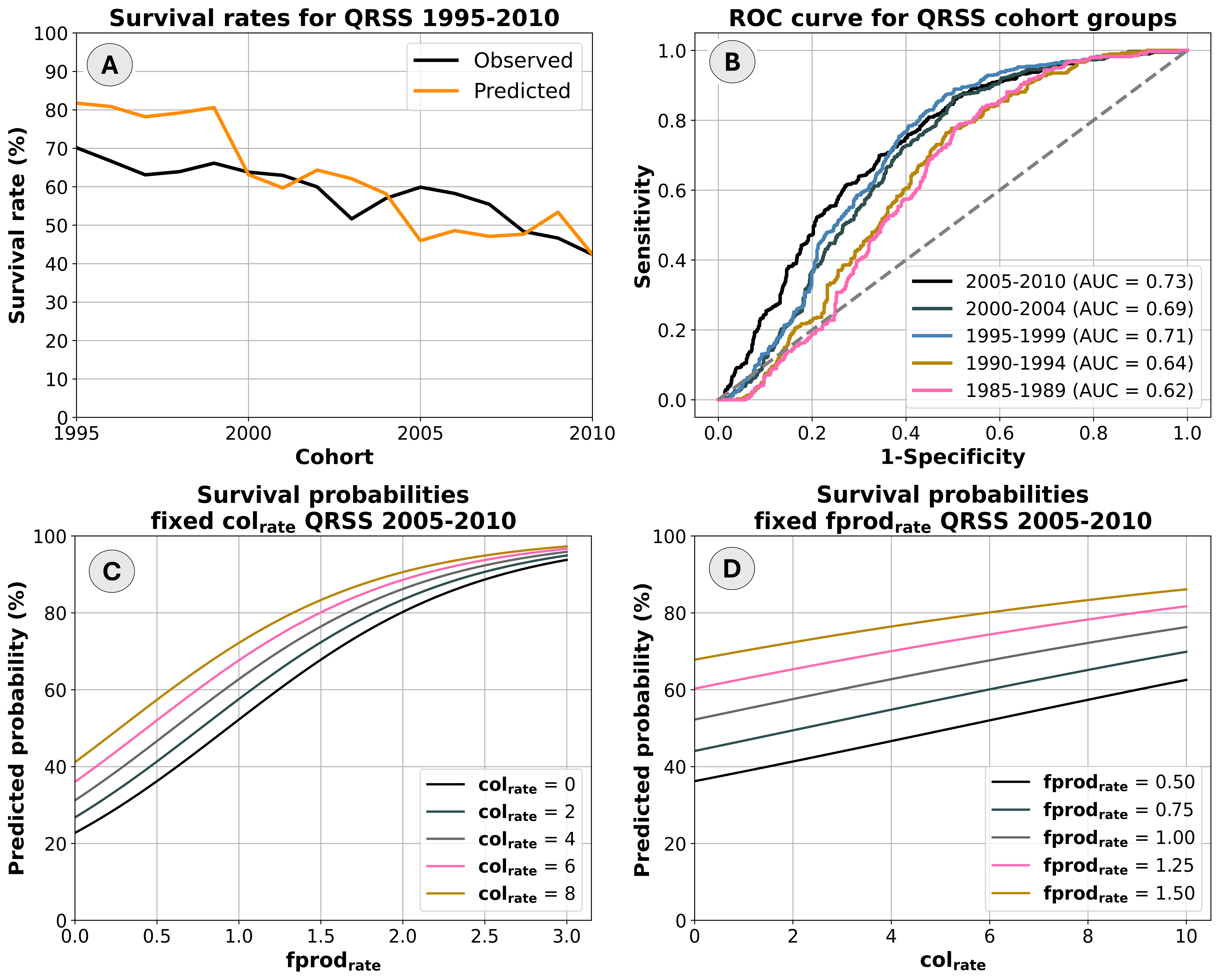}
    \caption{\small{Model performance of logistic regression with $\text{col}_{\text{\tiny{rate}}}$ and $\text{fprod}_{\text{\tiny{rate}}}$ as predictor variables. (A) Predicted and emperical survival curves for the QRSS 1995-2010. (B) ROC curves for logistic regressions on QRSS cohort groups. The grey dotted line indicates a random classifier. (C-D) Predicted survival probabilities on the QRSS 2005-2010 cohort group with one predictor held constant.}}
    \label{fig:modelperf}
\end{figure}

The 12-year survival curve in Fig. \ref{fig:surv_merged} for the QRSS 1995-2010 cohorts is estimated based on the average predicted survival ratios (using a classification threshold of 0.5). Results are shown in Fig. \ref{fig:modelperf} (top left). The model reproduces the salient trend of the empirical curve, with both over- and underestimation of survival rates at different times, while calibration improves from the 2000s onward. The ROC curves for the QRSS cohort groups are shown in Fig. \ref{fig:modelperf}  (top right).

In general performance of the two-factor model is moderate on all cohort groups from the 1995-2010. We note that the predictive abilities of the two-factor model analyzed in this section is not significant compared to the the two-factor models with $\text{fprod}_{\text{\tiny{rate}}}$ and $\text{cit}_{\text{\tiny{rate}}}$ or $\text{prod}_{\text{\tiny{rate}}}$ and $\text{cit}_{\text{\tiny{rate}}}$ as predictors (see Appendix \ref{multivariateanalysis}). In each case, however, performance likewise declines to levels comparable with a random classifier on older cohort groups. 
\section{Discussion} \label{discussion}


\noindent We have analyzed the career lengths of researchers in hep-th between 1950 and 2020. Our findings suggest a rapid transformation of the hep-th research community marked by changing demographics and increasing competition for survival. Our main observations are: 1) that the dropout ratios among junior researchers have increased significantly from the 1980s until the present, 2) surviving researchers have (on average) increased their collaboration and publication rates more than non-survivors during the same period, and 3) survivability can be predicted with moderate accuracy from the mid 1990s onwards using a simple two-factor regression model based on scientific proxies; this is not possible prior to the mid 1990s. Our results suggest that science proxies may have emerged as potential predictors of survivability in an increasingly competitive research environment. 

A central question is what incentive structures these developments may have created within the hep-th community. In this respect, we first note that contemporary science has been progressively influenced by ranking and assessment systems where, e.g., publication and citation counts are used to evaluate and reward academic performance \cite{Miller}. It is therefore natural to expect researchers to begin competing on these parameters. This is partially confirmed by our observation that mean early productivity rates are rising in hep-th, but that rates among survivors are higher than those of non-survivors with an increasing discrepancy in recent decades. This trend can, however, be explained by a simultaneous increase in the mean early collaboration rates among survivors compared to non-survivors. One possible consequence of stronger competition may be the incentive for authors to collaborate more as a way of enhancing their {\it full} productivity output. This effect will naturally disfavor researchers who work alone and may in part explain why this type of researcher is rapidly disappearing. A common alternative explanation (see for instance \cite{Wang_Barabási_2021,Giov} and references therein) is that collaboration rates have risen as means of travel and communication have become more easily accessible, or because it facilitates progress when working on complex problems. We note that this explanation does not exclude the one proposed here: That collaboration has become an increasingly viable survival strategy.

Regarding citations, we observe a difference among survivors and non-survivors in terms of mean citation rates, with survivors consistently achieving higher levels. The rates may, however, be impacted by cumulative advantages over lifetimes, i.e., that early papers by surviving authors are simply cited more often. In this respect we also note that the difference between survivors and non-survivors almost vanishes from the 1960s onwards when we consider mean citations per early paper. Assuming that citations reflect the impact or quality of research our results suggest that the early scientific output of hep-th dropouts is on par with those achieving full academic careers.

Furthermore, a key observation is that author survivability within the hep-th community can be modeled from the mid 1990s onward with moderate accuracy using a simple two-factor regression model based on early career collaborations and fractional productivity. Fitting the model with the same predictors on older cohorts leads to significant performance deterioration. A similar observation holds true for two-factor regression models with early citation rates and full or fractional productivity rates as predictors. In contrast to this, early citations per paper shows no significance in terms of predicting author survivability. The results indicate changing researcher behavior, with the hep-th community adapting in order to survive and becoming more homogenized. On a more general level, we note that using citations or productivity as measures when rewarding academic performance (e.g., by hiring) might create an incentive for researchers to maximize their measurement scores. In this regard, we refer to the principle of Goodhart’s law, which states that measures deteriorate when they become a target \cite{Fire, Mizen, Aknes}. Indeed, if selection pressures incentivize researchers to pursue higher citation counts or produce more papers (e.g., by collaborating more), then these measures will gradually reflect their ability to succeed in these parameters, rather than their innovation or the quality of their work. Furthermore, one possible way for researchers to gain collaborators, increase productivity, and improve their chances of being cited is to publish within popular areas of research and avoid working on controversial or niche ideas. Moreover, the advantage of having a large and well-connected research network is well-documented \cite{Raphael,Wang_Barabási_2021,Santo,LiW} and benefits primarily those working within mainstream research. Naturally, such tendencies will ultimately have a conservative, norm-preserving impact on the type of research being conducted, and potentially be detrimental to the progress of science itself. It may also partially explain the continued dominance of hep-th research programs that emerged in the 1970s and 1980s, which are still being pursued despite being challenged by empirical data, as well as the remarkable absence of conceptually new ideas within hep-th. 

Concerning sampling methods, assessing the level of competition in any research field will always be challenged by author migration. Since authors can move between different research topics over time, ensuring a homogeneous author population is difficult. We tried to alleviate this by sampling authors based on their early career research, excluding senior authors, who migrated to the same research field(s) at a later stage in their career. This method, however, does not consider authors who migrate away from the field. The survival rates obtained here therefore represent {\it academic} survival rates, rather than the actual survival rates in terms of remaining active within hep-th. These are different metrics, with the latter providing an upper bound of the former. We also note that the hep-th survival rates obtained for the QRSS and STQG samples in this paper are comparable to those obtained for the 2000 and 2010 physics cohorts in the study by Kwiek \& Szymula\cite{kwiek2024}, where authors were classified based on the journals of all cited papers in their lifetime publishing portfolio.

A central principle in our sampling was to prioritize data homogeneity rather than sample size. The use of very large samples without appropriate author or topic segmentation carries a risk obtaining spurious correlations or loss of information and accuracy due to noise build-up or systematic biases. Classifying authors according to their overall research topic (e.g, `physics') also implies sampling theoretical and experimental scientists from a variety of different fields, which may have different publication, collaboration, or citation practices. Furthermore, our results show a dramatic transformation of the hep-th field over the previous four decades, implying that author demographics from the 1980s are not representative of the present. Thus, evaluating potential lack of representativeness when working with samples covering long historical periods seems necessary. We note that this is at odds with numerous bibliometric studies that operate with large unsegmented samples or sample authors over long historical periods; see e.g.  \cite{Boothby,Zhao}.

The present study, however, is not without limitations. Refined segmentation will inevitably result in data reductions, which is a source of potential error in our analysis. The cohorts obtained here contain between 100 and 600 observations in most cases, with less than 2,000 observations for the most recent data even when grouped as in Table \ref{table:binning}. This naturally implies some uncertainty in our statistics. Data limitations may also impact our modeling, and we note that the performance of our logistic regression model only shows moderate discriminatory power, even on the most recent data, with relatively weak calibration. Considering the data limitations and the fact that the model only utilizes two predictor variables, we nonetheless assess that performance from the mid-1990s onward is significant. The observed performance deterioration is also noticeable, as the older cohort groups in the QED sample are larger than the cohort groups used to select the predictor variables. More sophisticated modeling techniques may also improve accuracy. 

We must also emphasize that many factors may influence researcher dropouts, e.g., a general lack of job satisfaction, family responsibilities, gender disparities, insufficient supervision or faculty support, and that the dropout ratios observed here cannot exclusively be attributed to academic competition \cite{Aldo,Dorenkamp,White,Hunter,Maher}. In general, many confounding factors may impact hiring and selection procedures, which are not necessarily reflected in bibliometric data.


Finally, we must remark on the general lack of cohort-based or longitudinal studies of researcher survival rates \cite{Milojevic, kwiek2024, Boothby}. There can be no doubt that competition for survival is one of the most important sociological factors in contemporary academia, and thus it is surprising that its impact on science have not been more thoroughly analyzed. We hope that this paper will stimulate further research in this direction.

\section*{Acknowledgements} JS is indebted to Linus Emborg for many inspiring discussions and helpful suggestions. 
JMG would like to express his gratitude to entrepreneurs Kasper Bloch Gevaldig and Jeff Cordova for their generous financial support. JMG is also indebted to the following list of sponsors for their support:   Frank Jumppanen Andersen, Danny Halskov Birkmose, Bart De Boeck, Simon Chislett, Jos van Egmond, Trevor Elkington, Jos Gubbels, Claus Hansen, David Hershberger, Ilyas Khan, Simon Kitson, Hans-J\o rgen Mogensen, Stephan M{\"u}hlstrasser, Bert Petersen, Ben Tesch, Adam Tombleson, Jeppe Trautner, Vladimir Zakharov, and the company Providential Stuff LLC. 




\appendix




\section{Data corrections and exclusions} \label{datacorr}
Search filters used in our initial OpenAlex query are shown in Table \ref{table:OpenAlex}.
\begin{table}[H]
\centering
\scalebox{0.8}{\begin{tabular}{l||l|l|l|l}
\hline
\textbf{Sample name (x)} & \textbf{Year} & \textbf{Concept(s)} & \textbf{Source} & \textbf{Source type} \\
\hline
\hline
QED & 1950-1990 & ``Quantum electrodynamics'' & Article & Journal \\
 \hline
QRSS & 1970-2010 & \begin{tabular}[x]{@{}l@{}}``Quantum field theory''\\``Renormalization group''\\``Supersymmetry''\end{tabular} & Article & Journal\\
\hline
STQG & 1980-2020  &  \begin{tabular}[x]{@{}l@{}} ``String theory''\\ ``Quantum gravity''\end{tabular} & Article & Journal \\
\hline
\end{tabular}
}
\caption{\small{Seach filters in OpenAlex to generate the raw publication samples $\mathcal{D}_{\tiny{x}}$, $ x \in \lbrace \textnormal{QED, QRSS, STQG} \rbrace$.}}
\label{table:OpenAlex}
\end{table}
\noindent
To ensure a sample of hep-th papers only, the following exclusions were applied to $\mathcal{D}_{\tiny{x}},  x \in \lbrace \textnormal{QED, QRSS, STQG} \rbrace$:
\begin{enumerate}
    \item[-]\textit{Dubious publications.} Papers with dubious strings in their titles (e.g., `exercises', `introduction', `monetary', `policies', `book review') were excluded. Additionally, articles published in journals with dubious names (e.g., `teaching', `econometrics', `geology', `computer') were excluded.
    \item[-]\textit{Rare journals.} Journals appearing 5 times or less in the entire sample were excluded, as they were assessed to be mis-classifications by OpenAlex.
\end{enumerate}
The associated data reductions are shown in Table \ref{tab:corrections1}.
\begin{table}[H]
    \centering
    \scalebox{0.8}{\begin{tabular}{ p{4cm}|p{2cm}|p{2cm}|p{2cm} }
 \hline
 \multicolumn{4}{c}{\textbf{Reductions of $\mathcal{D}_{\text{\tiny{x}}}$}} \\
 \hline
 \hline
 \textbf{Step} & $\mathcal{D}_{\text{\tiny{QED}}}$ & $\mathcal{D}_{\text{\tiny{QRSS}}}$ & $\mathcal{D}_{\text{\tiny{STQG}}}$ \\
 \hline
 RAW   & 79,520 & 55,897 & 42,190 \\
 Dubious publications & 79,216 & 55,500 & 41,870 \\
 Rare journals & 77,983 & 52,563  & 39,374  \\
  \hline
  \hline
 \textbf{Total reduction} & 1,537 (1.93\%) & 3,334 (5.96\%)  & 2,816 (6.67\%)  \\
  \hline
\end{tabular}}
    \caption{\small{Corrections and exclusions performed on the raw publication hep-th publication samples, with the total number of publications after each data step shown.}}
    \label{tab:corrections1}
\end{table}
The set of unique authors $\mathcal{A}_x$ in $\mathcal{D}_x$ were obtained and their full career publications $\mathcal{D}_{x,\text{\tiny full}}$ were extracted from OpenAlex. The following corrections and exclusions steps were then performed on $\mathcal{D}_{x,\text{\tiny full}}$:
\begin{enumerate}
    \item[-]\textit{Missing values and duplicates.} All papers with missing titles, journal names or publication years were removed. Additionally, all duplicate papers with overlapping titles and first author ID were removed. Note that the data step removing duplicates resulted in a significant reduction as the same papers appeared for co-authors. 
    \item[-]\textit{Dubious years.} Pre-WWII publications were removed. This was both an intermediate data step, as authors with publications prior to the historical period defined in our scope would be removed regardless, but also removed possible mis-classification of publications by authors active within the period.
    \item[-]\textit{Many-author papers.} Papers with 10 authors or more were excluded from the data. We assessed that hep-th articles with 10 authors or more would be extremely rare. A manual inspection of the exclusions indicated that a significant part of the excluded papers were either not original research articles within hep-th but progress reports or monographs etc., or articles within applied sciences or other academic fields. The former should not be counted as publications in our method and the latter were excluded to obtain more accurate lifetimes for authors who did not remain active within the hep-th community. 
\end{enumerate}
The associated data reductions are shown in Table \ref{tab:corrections2}. 
\begin{table}[H]
    \centering
    \scalebox{0.8}{\begin{tabular}{ p{4cm}|p{2cm}|p{2cm}|p{2cm} }
 \hline
 \multicolumn{4}{c}{\textbf{Reductions of $\mathcal{D}_{\text{\tiny{x, full}}}$}} \\
 \hline
 \hline
 \textbf{Step} & $\mathcal{D}_{\text{\tiny{QED, full}}}$ & $\mathcal{D}_{\text{\tiny{QRSS, full}}}$ & $\mathcal{D}_{\text{\tiny{STQG, full}}}$ \\
 \hline
 RAW   & 4,394,065 & 4,208,918 & 2,154,668 \\
 Missings and duplicates & 3,319,619 & 2,086,045 & 1,302,824 \\
 Dubious years & 3,280,706 & 2,053,891 & 1,301,526  \\
 Many-author papers & 2,999,342 & 1,838,065 & 1,170,214  \\
  \hline
  \hline
 \textbf{Total reduction} & 1,394,723 (31.7\%) & 2,370,195 (56.3\%) & 984,454 (45.7\%)  \\
  \hline
\end{tabular}}
    \caption{\small{Corrections and exclusions performed on the career full publication hep-th samples, with the total number of publications after each data step shown.}}
    \label{tab:corrections2}
\end{table}
All metrics (lifetimes, early publications, early collaborators) of the authors in 
$\mathcal{A}_{\tiny{x}}$ were then calculated based on the publication data in $\mathcal{D}_{\tiny{x,\text{full}}}$. The following exclusions in $\mathcal{A}_{\tiny{x}}$ were then performed:
\begin{enumerate}
    \item[-]\textit{Authors with more than 500 publications.} Authors with more than 500 publications in $\mathcal{D}_{\tiny{x,\text{full}}}$ were excluded, as they were assessed to have inflated publication numbers in OpenAlex. 
    \item[-]\textit{Low productivity authors.} Authors with extremely low productivity rates $\left(\frac{\text{total publications}}{\text{total lifetime}} < \frac{1}{3}\right)$ in $\mathcal{D}_{\tiny{x,\text{full}}}$ were excluded, as they were assessed to have spurious publication data in OpenAlex, distinguished by very large gaps in their publication history. These authors would in turn have inflated lifetimes and induce a bias in the statistics. The issue was most pervasive in the QED sample, which contain the oldest data.    
    \item[-]\textit{Authors with early first publications.} Authors whose first publication in $\mathcal{D}_{\tiny{x,\text{full}}}$ came decades before the historical period within our scope were removed. This was an intermediate data step. 
    \item[-]\textit{Authors not in} $\mathcal{D}_{\tiny{x}} \cap \mathcal{D}_{\tiny{x,\text{full}}}$. Authors with no publications in $\mathcal{D}_{\tiny{x}} \cap \mathcal{D}_{\tiny{x,\text{full}}}$ after the above exclusions were removed.
    \item[-]\textit{Authors where $\Delta > 3$}. Authors whose first publication in $\mathcal{D}_{\tiny{x}}$ came later than 3 years after their first publication in $\mathcal{D}_{\tiny{x,\text{full}}}$ were excluded as per our scope. Note that this resulted in the most significant data reduction.
    \item[-]\textit{Transient authors}. All transient authors, i.e., authors with $T(r)=1$, were excluded.
    \item[-]\textit{Final exclusions}. Finally, the samples were reduced to the historical periods within our scope (see Table \ref{table:Samples}). Additionally, outlier analysis of key variables identified a small number of authors with erroneous publication, citation and collaboration numbers, confirmed by manual inspection, resulting in minor exclusions ($<20$) for each sample. 
\end{enumerate}
The associated data reductions are shown in Table \ref{tab:corrections3}.
\begin{table}[H]
    \centering
    \scalebox{0.8}{\begin{tabular}{ p{5cm}|p{2cm}|p{2cm}|p{2cm} }
 \hline
 \multicolumn{4}{c}{\textbf{Reductions of $\mathcal{A}_{\tiny{x}}$}} \\
 \hline
 \hline
 \textbf{Step} & $\mathcal{A}_{\text{\tiny{QED}}}$ & $\mathcal{A}_{\text{\tiny{QRSS}}}$ & $\mathcal{A}_{\text{\tiny{STQG}}}$ \\
 \hline
RAW & 47,758 & 29,207 & 21,044 \\
More than 500 publications & 47,232 & 28,880 & 20,854 \\
$P_{\tiny{\text{tot}}}/T_{\tiny{\text{tot}}} < 1/3$ & 43,131 & 27,221 & 20,055  \\
Early first publications & 43,131 & 26,906 & 19,769  \\
Not in $\mathcal{D}_{\tiny{x}} \cap \mathcal{D}_{\tiny{x,\text{full}}}$ & 41,673 & 24,087 & 18,696 \\
$\Delta > 3$ & 22,543 & 10,369 & 7,573 \\
Transients & 18,490 & 8,906 & 6,578 \\
Final exclusions & 15,356 & 8,570 & 6,223  \\
  \hline
  \hline
 \textbf{Total reduction} & 32,399 (67.8\%)  & 20,637 (70.7\%) & 14,820 (70.4\%)  \\
  \hline
\end{tabular}}
    \caption{\small{Corrections and exclusions performed on the hep-th author samples, showing the total number of authors after each data step.}}
    \label{tab:corrections3}
\end{table}
\section{Cohort sizes} \label{cohortsizes}
\begin{figure}[H] 
    \centering
    \includegraphics[width=1\linewidth]{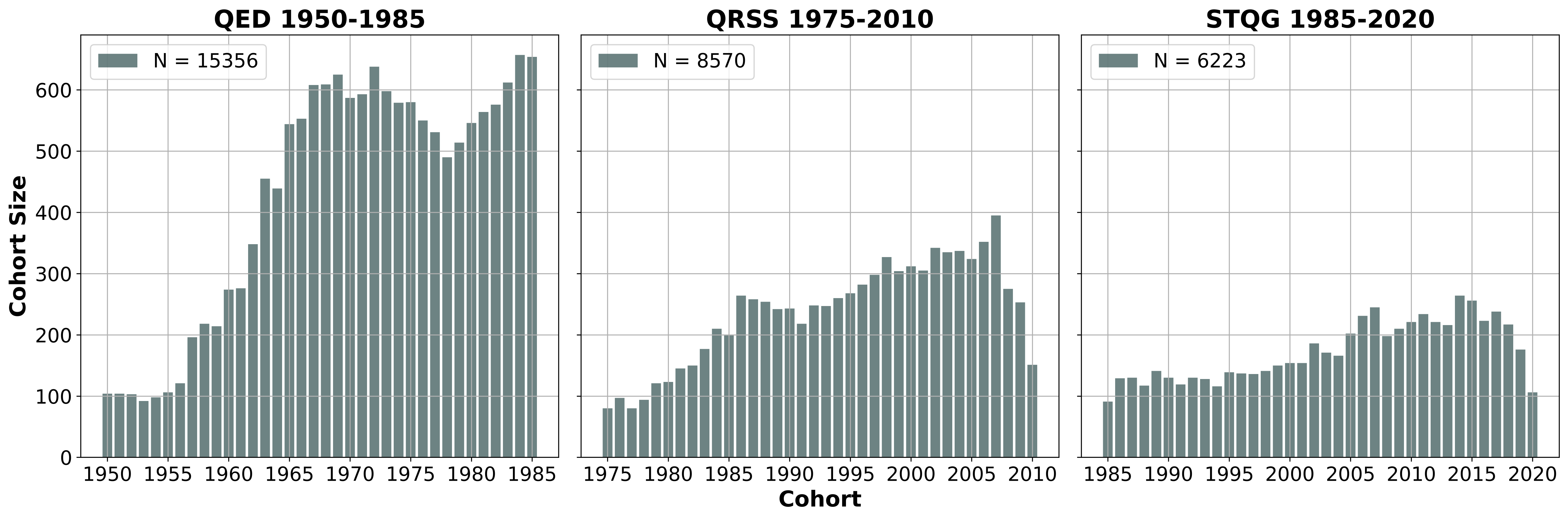}
    \caption{\small{Cohort sizes. The total number of authors in each cohort for the QED (left), QRSS (middle) and STQG (right) sample.}}
    \label{fig:cohortsizes}
\end{figure}

After all data exclusions and corrections, authors in $\mathcal{A}_{\tiny{x}}$ were grouped into cohorts based on their initial publication year. The cohort sizes for each sample is shown in Fig. \ref{fig:cohortsizes}.

\section{Variable Distributions}\label{vardistributions}
\begin{figure}[H] 
    \centering
    \includegraphics[width=1\linewidth]{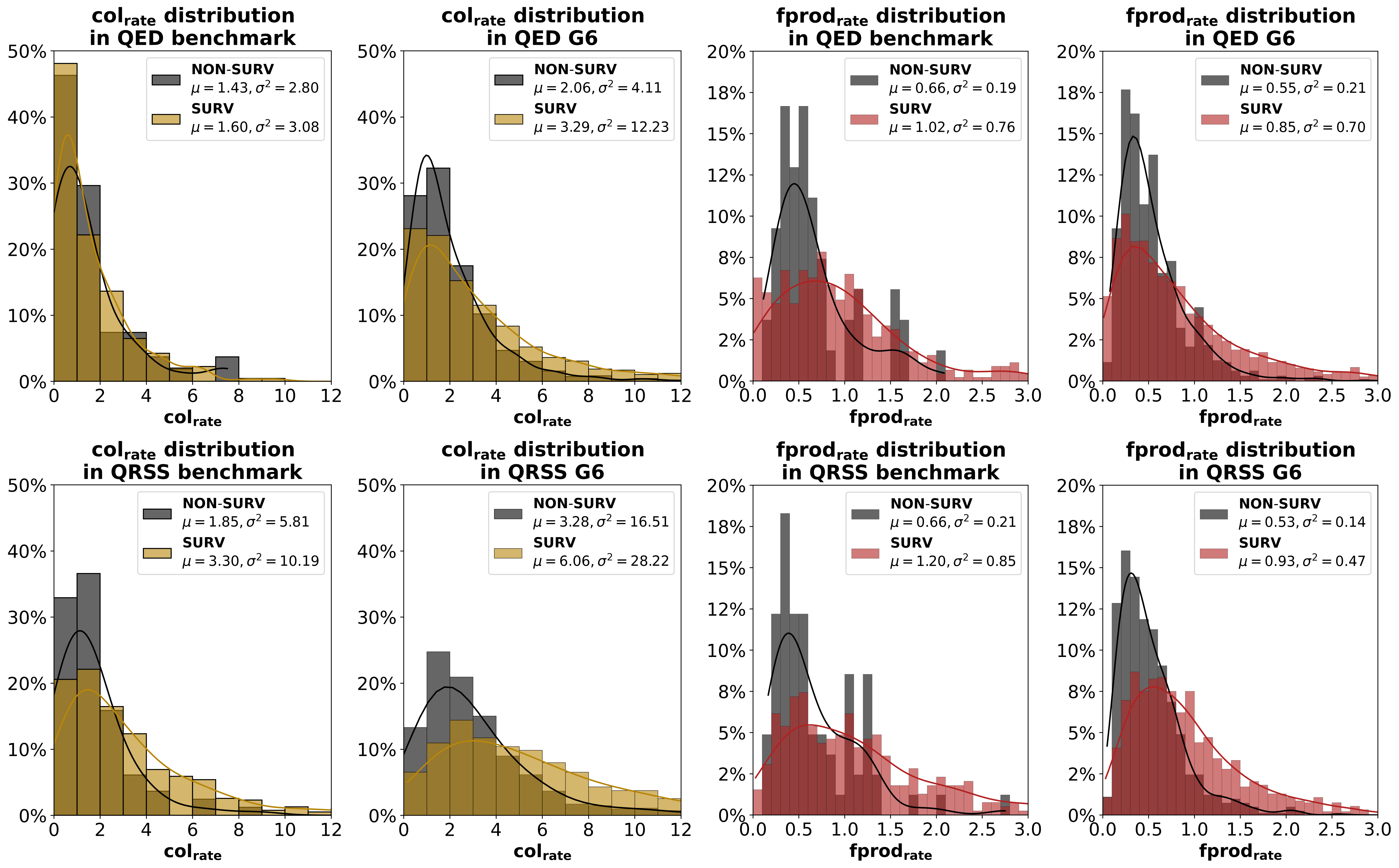}
    \caption{\small{Variable distributions. PMF (probability mass function) with KDE (kernel density estimation) for $\text{col}_{{\text{\tiny{rate}}}}$ and $\text{fprod}_{{\text{\tiny{rate}}}}$ among survivors and non-survivors in the benchmark and G6 cohort groups for the QED (top) and QRSS (bottom) samples.}}
    \label{fig:dist_merge}
\end{figure}

\noindent Fig. \ref{fig:dist_merge} shows the distributions for $\text{col}_{{\text{\tiny{rate}}}}$ and $\text{fprod}_{{\text{\tiny{rate}}}}$ among authors in the benchmark and G6 cohort groups for the QED and QRSS samples. 
The corresponding distributions for the benchmark and G4 groups in the STQG sample are shown in Fig. \ref{fig:PSIcorr} (middle).
\section{Univariate analysis}\label{univariateanalysis}
\noindent
Table \ref{table:univaranalysis} summarizes the results of the univariate analysis from the step-forward selection procedure. All variables except $\text{C}_{\text{\tiny{p}}}\text{P}$ passed the retention criteria.
\begin{table}[H]
\centering
\scalebox{0.8}{
\begin{tabular}[t]{l|lccc}
\hline
\textbf{Cohort group} & Predictor & AUROC & $z$ & $p>|z|$  \\
\hline
\hline
& $\text{col}_{\text{\tiny{rate}}}$& 0.68 & -7.82850 & $<0.001$ \\
& $\text{prod}_{\text{\tiny{rate}}}$& 0.72 & -10.0672 & $<0.001$ \\
STQG G4 & $\text{fprod}_{\text{\tiny{rate}}}$& 0.71 & -9.68699 & $<0.001$ \\
& $\text{cit}_{\text{\tiny{rate}}}$& 0.69 & -8.22688 & $<0.001$ \\
& $\text{C}_{\tiny{p}}\text{P} $& 0.60  & -2.57779 & $0.0184$ \\
\hline 
& $\text{col}_{\text{\tiny{rate}}}$& 0.70 & -10.281 & $<0.001$ \\
& $\text{prod}_{\text{\tiny{rate}}}$& 0.73 & -12.561 & $<0.001$ \\
QRSS G6 & $\text{fprod}_{\text{\tiny{rate}}}$& 0.71 & 11.809 & $<0.001$ \\
& $\text{cit}_{\text{\tiny{rate}}}$& 0.71 & -10.105 & $<0.001$ \\
& $\text{C}_{\text{\tiny{p}}}\text{P}$ & 0.63 & -5.06784 & $<0.001$ \\
\hline 
\end{tabular}
}
\caption{\small{Univariate Analysis. Mean results of univariate logistic regression with key variables as predictors, fitted on cohort group G4 (2005-2010) in the STQG sample and cohort group G6 (2005-2010) in the QRSS sample over 10,000 iterations.}}
\label{table:univaranalysis}
\end{table}
\section{Multivariate analysis}\label{multivariateanalysis}
\noindent 
This section summarizes the performance of the multivariate models in our step-forward selection procedure. All models were fitted on the STQG G4 (2005-2010) and QRSS G6 (2005-2010) cohort groups over 10,000 iterations. Additionally, models with acceptable discriminatory power and significant variables were tested on older cohort groups to assess if the performance deterioration observed in the final model was pervasive. Table \ref{table:alternativemodel1} shows the mean results for the two-factor model with $\text{prod}_{\text{\tiny{rate}}}$ and $\text{cit}_{\text{\tiny{rate}}}$ as predictors, and Table \ref{table:alternativemodel2} the two-factor model with $\text{fprod}_{\text{\tiny{rate}}}$ and $\text{prod}_{\text{\tiny{rate}}}$ as predictors. Average performance of these model candidates were on par with the final model using $\text{col}_{\text{\tiny{rate}}}$ and $\text{fprod}_{\text{\tiny{rate}}}$ as predictors. Results for the final model are shown in Table \ref{table:finalmodel} for comparison. All other multivariate models were discarded due to low discriminatory power, lack of predictor significance or high uncertainty in the parameter estimates. 
\begin{table}[H]
\centering
\scalebox{0.8}{
\begin{tabular}[t]{l|clcc}
\hline
\textbf{Cohort group} & AUROC & Predictor & $z$ & $p>|z|$  \\
\hline
\hline
QED G5 & 0.58  & $\text{prod}_{\text{\tiny{rate}}}$  & -4.49806  & $<0.001$ \\
(1975-1979) & & $\text{cit}_{\text{\tiny{rate}}}$  &  -2.47944 & $0.0184$  \\
\hline
QED G6 & 0.61  & $\text{prod}_{\text{\tiny{rate}}}$  & -7.50862 & $<0.001$ \\
(1980-1985) & & $\text{cit}_{\text{\tiny{rate}}}$  & -2.56863 & $0.0249$   \\
\hline
QRSS G2 & 0.64 & $\text{prod}_{\text{\tiny{rate}}}$  & -4.19591 & $<0.001$ \\
(1985-1989) & & $\text{cit}_{\text{\tiny{rate}}}$  & -2.39088 & $0.0240$  \\
\hline
QRSS G3 & 0.62 & $\text{prod}_{\text{\tiny{rate}}}$  & -5.64243 & $<0.001$ \\
(1990-1994) & & $\text{cit}_{\text{\tiny{rate}}}$  & -0.75252 & $0.0473$  \\
\hline
QRSS G4 & 0.70 & $\text{prod}_{\text{\tiny{rate}}}$  & -7.87137 & $<0.001$ \\
(1995-1999) & & $\text{cit}_{\text{\tiny{rate}}}$  & -1.90337 & $0.0819$  \\
\hline
QRSS G5 & 0.68 & $\text{prod}_{\text{\tiny{rate}}}$  &  -9.37868 & $<0.001$ \\
(2000-2004) & & $\text{cit}_{\text{\tiny{rate}}}$  & -0.81295 & $0.0596$  \\
\hline
QRSS G6 & 0.73 & $\text{prod}_{\text{\tiny{rate}}}$  &  -8.12771 & $<0.001$ \\
(2005-2010) & & $\text{cit}_{\text{\tiny{rate}}}$  & -2.90010 & $0.0159$  \\
\hline 
STQG G4 & 0.72 & $\text{prod}_{\text{\tiny{rate}}}$  &  -7.33140 & $<0.001$ \\
(2005-2010) & & $\text{cit}_{\text{\tiny{rate}}}$  & -3.34488 & $0.0027$  \\
\hline 
\end{tabular}
}
\caption{\small{Average performance over 10,000 iterations of the two-factor logistic regression model with $\textnormal{prod}_{\textnormal{\tiny{rate}}}$ and $\textnormal{cit}_{\textnormal{\tiny{rate}}}$ as predictors.}}
\label{table:alternativemodel1}
\end{table}
\begin{table}[H]
\centering
\scalebox{0.8}{
\begin{tabular}[t]{l|clcc}
\hline
\textbf{Cohort group} & AUROC & Predictor & $z$ & $p>|z|$  \\
\hline
\hline
QED G5 & 0.58 & $\text{fprod}_{\text{\tiny{rate}}}$  & -4.22705  & $<0.001$  \\
(1975-1979) & & $\text{cit}_{\text{\tiny{rate}}}$  &  -2.82213 & $0.0075$  \\
\hline
QED G6 & 0.61 & $\text{fprod}_{\text{\tiny{rate}}}$  & -6.33860 & $<0.001$ \\
(1980-1985) & & $\text{cit}_{\text{\tiny{rate}}}$  & -3.73640 & $<0.001$   \\
\hline
QRSS G2 & 0.63 & $\text{fprod}_{\text{\tiny{rate}}}$  & -3.71618 & $<0.001$ \\
(1985-1989) & & $\text{cit}_{\text{\tiny{rate}}}$  & -2.98896 & $0.0047$  \\
\hline
QRSS G3 & 0.61 & $\text{fprod}_{\text{\tiny{rate}}}$  & -4.54405 & $<0.001$ \\
(1990-1994) & & $\text{cit}_{\text{\tiny{rate}}}$  & -1.87205 & $0.0769$  \\
\hline
QRSS G4 & 0.71 & $\text{fprod}_{\text{\tiny{rate}}}$  & -7.04799 & $<0.001$ \\
(1995-1999) & & $\text{cit}_{\text{\tiny{rate}}}$  & -3.07670 & $0.0048$  \\
\hline
QRSS G5 & 0.67 & $\text{fprod}_{\text{\tiny{rate}}}$  &  -7.23875 & $<0.001$ \\
(2000-2004) & & $\text{cit}_{\text{\tiny{rate}}}$  & -2.67829 & $0.0277$  \\
\hline
QRSS G6 & 0.72 & $\text{fprod}_{\text{\tiny{rate}}}$  &  -7.40857 & $<0.001$ \\
(2005-2010) & & $\text{cit}_{\text{\tiny{rate}}}$  & -5.20633 & $<0.001$  \\
\hline 
STQG G4 & 0.72 & $\text{fprod}_{\text{\tiny{rate}}}$  &  -6.59763 & $<0.001$ \\
(2005-2010) & & $\text{cit}_{\text{\tiny{rate}}}$  & -4.44598 & $<0.001$ \\
\hline 
\end{tabular}
}
\caption{\small{Average performance over 10,000 iterations of the two-factor logistic regression model with $\textnormal{fprod}_{\textnormal{\tiny{rate}}}$ and $\textnormal{cit}_{\textnormal{\tiny{rate}}}$ as predictors.}}
\label{table:alternativemodel2}
\end{table}
\begin{table}[H]
\centering
\scalebox{0.8}{
\begin{tabular}[t]{l|clcc}
\hline
\textbf{Cohort group} & AUROC & Predictor & $z$ & $p>|z|$  \\
\hline
\hline
QED G5 & 0.59 & $\text{fprod}_{\text{\tiny{rate}}}$  & $-5.03106$ & $<0.001$\\
(1975-1979) & & $\text{col}_{\text{\tiny{rate}}}$  & $-3.08188$  &  $0.0042$ \\
\hline
QED G6 & 0.62 & $\text{fprod}_{\text{\tiny{rate}}}$  & $-6.52902$ & $<0.001$  \\
(1980-1985) & & $\text{col}_{\text{\tiny{rate}}}$  & $-5.35400$ &  $<0.001$ \\
\hline
QRSS G2 & 0.62 & $\text{fprod}_{\text{\tiny{rate}}}$  & -2.92415 & $<0.001$ \\
(1985-1989) & & $\text{col}_{\text{\tiny{rate}}}$  & -4.01193 & $0.0072$  \\
\hline
QRSS G3 & 0.63 & $\text{fprod}_{\text{\tiny{rate}}}$  & -4.06742 & $<0.001$ \\
(1990-1994) & & $\text{col}_{\text{\tiny{rate}}}$  & -3.9769 & $0.0769$  \\
\hline
QRSS G4 & 0.71 & $\text{fprod}_{\text{\tiny{rate}}}$  & -4.446537 & $<0.001$ \\
(1995-1999) & & $\text{col}_{\text{\tiny{rate}}}$  & -6.75604 & $0.0048$  \\
\hline
QRSS G5 & 0.69 & $\text{fprod}_{\text{\tiny{rate}}}$  &  -6.16488 & $<0.001$ \\
(2000-2004) & & $\text{col}_{\text{\tiny{rate}}}$  & -5.94704 & $0.0277$  \\
\hline
QRSS G6 & 0.73 & $\text{fprod}_{\text{\tiny{rate}}}$  &  -8.20081 & $<0.001$ \\
(2005-2010) & & $\text{col}_{\text{\tiny{rate}}}$  & -5.28358 & $<0.001$  \\
\hline 
STQG G4 & 0.72 & $\text{fprod}_{\text{\tiny{rate}}}$  & -7.33141 & $<0.001$ \\
(2005-2010) & & $\text{col}_{\text{\tiny{rate}}}$  & -3.3448 & $0.0026$ \\
\hline 
\end{tabular}
}
\caption{\small{Average performance over 10,000 iterations of the two-factor logistic regression model with $\textnormal{col}_{\textnormal{\tiny{rate}}}$ and $\textnormal{fprod}_{\textnormal{\tiny{rate}}}$ as predictors.}}
\label{table:finalmodel}
\end{table}

\medskip



\begin{thebibliography}{99}
\bibitem{Scheidsteger} T. Scheidsteger and R. Haunschild, ``{Which of the metadata with relevance for bibliometrics are the same and which are different when switching from Microsoft Academic Graph to OpenAlex?}", Profesional De La Información, \textbf{32} (2), (2023).
\end{thebibliography}
\end{document}